\documentclass[twocolumn,showpacs,preprintnumbers,amsmath,amssymb]{revtex4}
\begin{document}

\preprint{}

\title{Directed random walks on hierarchic trees with continuous branching: a renormalization group approach}

\author{David B. Saakian$^{1,2,3}$}
\email{saakian@yerphi.am}

\affiliation{$^1$Institute of Physics, Academia Sinica, Nankang,
Taipei 11529, Taiwan} \affiliation{$^2$Yerevan Physics Institute,2
Alikhanian Brothers St., Yerevan 375036, Armenia}
\affiliation{$^3$National Center for Theoretical Sciences
(North):Physics Division, National Taiwan University, Taipei 10617,
Taiwan}

\begin{abstract}
We investigate the directed  random walk on hierarchic trees. Two
cases are investigated: random variables on deterministic trees with
a continuous branching, and random variables on the trees
constructed trough the random branching process. We derive
renormalization group (partial differential) equations for the
branching models with binomial, Poisson and compound Poisson
distributions of random variables on the links of tree. These
renormaliation group equations are new class of reaction-diffusion
equations in 1-dimension.
\end{abstract}
\pacs{05.40.-a, 11.25.Mj, 75.10.Nr}
\maketitle

\section{Introduction}
\label{introduction}

Models on hierarchic trees are rather popular in statistical
physics \cite{ba}. Hierarchic tree models with random quenched
disorder \cite{de88}-\cite{de93} are especially popular object of
research, they are connected with Random Energy Model. Due to
special geometry, these models could be solved trough recursive
equations \cite{ba}. Considering random walks (directed polymers)
on trees, the recursive equation for these models have been
approximated with Kolmogorov- Petrovsky-Piscounov (KPP)equation.

The investigation of statistical physics models with quenched
disorder on hierarchic trees with continuous branching   are the
subject of extensive studies. The models with continuous branching
are especially popular in financial market literature
\cite{cf02},\cite{mu06}, as well as in turbulence. In \cite{sa02}
has been considered hierarchic trees with continuous branching,
when there are random variable with normal distribution on the
tree.  These version of models later has been investigated in
\cite{sa09,sa10},\cite{fy09}, applying to the problems of
disordered systems and string theory. To solve hierarchic models
with continuous branching, in \cite{sa02} has been used a
renormalization group equation, first obtained in \cite{do}. This
equation is similar to KPP equation, see \cite{fy09} for
discussion.
 In the current article we will derive renormalization
group equation for the general case of distribution for random
variables on the branches of hierarchic tree.

We are going to investigate the random directed walks on
hierarchic trees described trough partition function:
\begin{eqnarray}
\label{e1} Z=\sum_{i}\exp[-\beta y_i]
\end{eqnarray}
where $y_i$ are defined on the endpoints of the tree and $\beta$ is
the inverse temperature.

The model is defined by considering a tree with branching number q
and K levels of hierarchy, therefore there are $q^K\equiv e^L$
endpoints of the tree. At any level of hierarchy there are q links
from every node, and on every link a random variable $\epsilon_l$ is
defined. The variable $y_i$ entering the definition of the partition
function Eq.(1) is associated with a unique path going from the
origin O to the i-th tree endpoint and amounts to summing up
variables $\epsilon_l$ along this path,
\begin{eqnarray}
\label{e2} y_i=\sum_l\epsilon_l
\end{eqnarray}

One can calculate mean free energy $\ln Z$ using the identity
\cite{de80}
\begin{eqnarray}
\label{e3} <ln
Z>=\int_0^{\infty}(\frac{e^{-p}}{p}-\frac{<e^{-pZ}>}{p})dp=\nonumber\\
\Gamma'(1)+\int_0^{\infty}\ln p d<e^{-t Z}>
\end{eqnarray}
Thus  we need to calculate
\begin{eqnarray}
\label{e4} g(p,\beta)\equiv G_K(x)=<\exp[-p\sum_ie^{\beta
y_i}]>|_{\epsilon_l},\nonumber\\
p=e^{-\beta x}
\end{eqnarray}
for the average $<\exp[-p\sum_ie^{-\beta y_i}]>|_{\epsilon_l}$ over
the configurations on the K-level hierarchic tree.

Having the expression for $g(p,\beta)$, we can calculate the moments
of $Z^{-n}$:
\begin{eqnarray}
\label{e5} <Z^{-n}>=\int_0^{\infty}dpp^ng(p,\beta)
\end{eqnarray}

 In Ref.~\cite{de88} an equivalent
problem has been considered, and recursive relations have been
introduced . Following \cite{de91}, we define \begin{eqnarray}
\label{e6} G_0(x)=\exp[-e^{-\beta x}]
\end{eqnarray}
and then other $K$ functions $G_l(x)=<\exp[-e^{-\beta
x}\sum_ie^{-\beta y_i}]>|_{\epsilon_l}$ for the model on the trees
with $l$ level of hierarchy, $0\le l\le K$. We identify
$g(p,\beta)\equiv G_K(-\ln p/\beta)$.

 Consider the l-level
hierarchic tree. It could be fractured into q
 trees with hierarchy level $l-1$ each. The $y_i$ at the endpoints of
 l-level tree can be derived from the $y_i$
of the $(l-1)$ level tree adding a random variable $\epsilon$.
  A simple consideration gives a recursive relation \cite{de88,de91}:
\begin{eqnarray}
\label{e7} G_{l}(x)=[\int d\epsilon\;
G_{l-1}(x+\epsilon)\rho(\epsilon)]^q
\end{eqnarray}
Considering recursive relations for $1\le l\le K$ with the boundary
condition by Eq. (6), we can calculate $G_K$.

Eq.(7), derived in \cite{de88}-\cite{de91} is an exact recursive
relation. Similar recursive equations have been considered in the
research of real space renormalization approach to quantum disorder
in d-dimensional space, see Eq.(32) in \cite{ga10}. The latter is
some approximation, while Eq.(7) is an exact relation. The more
serious difference- our random variables $\epsilon$ have independent
distributions for different hierarchy levels {\it l}, while at
quantum disorder point [20] there is some correlation of noise at
different hierarchy levels. To get some analytical estimates, in
\cite{ga10} has been considered the limit of large branching number
$q$ (K in notation of \cite{ga10}), similar to those used in
\cite{de91}. \cite{ga10} used general distribution of random
variables.

The KPP like equation has been applied for investigation of quenched
disorder model. In the next section we will consider the $q\to 1$
limit of Eq.(7), deriving renormalization group (KPP like) equation
for the general distribution of $\epsilon$.

\section{Random walks on a hierarchic tree}

\subsection{Renormalization-group equation for Hierarchic trees with continuous branching}

While Eq.(7) is derived for integer q, we can consider the
equation for any positive value of q as well. Let us consider the
model, where the branching number is close to $1$.
 We can now derive exact differential equation instead of iteration equation Eq.~(\ref{e7}).
 We have $K=1/\Delta v$ levels of branching on our tree with $q=e^{L\Delta
 v}$, where $L\Delta v\ll 1$.
 We identify
\begin{eqnarray}
\label{e8} G_l(x)\equiv G(x,v),\nonumber\\
v=\frac{l}{K},\nonumber\\
g(p,\beta)=G(-\frac{\ln p}{\beta})
\end{eqnarray}
To define a distribution for random variable $\epsilon_l$,  let us
start with some random distribution $\hat \rho(x)$. In the next step
we
 calculate the distribution for the sum of $L$
random variables with such distributions. For our purposes it is
convenient to use the following representation for $\hat \rho$:
\begin{eqnarray}
\label{e9} \hat\rho(x)=\frac{1}{2\pi}\;\int_{-\infty}^{\infty}\; dh
\;\exp[\phi(ih)-ihx]
\end{eqnarray}
Then for the $\epsilon$, a sum of L random variables x, we get the
compose distribution $\rho(L,\epsilon)$ just multiplying $\phi$ in
the exponent 0f Eq. (9) by L:
\begin{eqnarray}
\label{e10}
\rho(L,\epsilon)=\frac{1}{2\pi}\;\int_{-\infty}^{\infty}\; dh
\;\exp[L\phi(ih)-ih\epsilon]
\end{eqnarray}
To ensure the probability balance condition, there is a constraint
\begin{eqnarray}
\label{e11}\phi(0)=0
\end{eqnarray}
We can consider the Taylor expansion
\begin{eqnarray}
\label{e12}\phi(ik)=\sum_{l\ge 1}b_l(ik)^l
\end{eqnarray}

According to our notations $\hat \rho(x)\equiv \rho(1,x)$. A
concrete case of the latter distribution with $\ln cosh(h)$ has been
used while considering the diluted REM \cite{sa97}. Here we will
consider another distributions, popular in turbulence and financial
theory.

We take $\hat \rho(L,y_i)$ as a distribution of $y_i$. The L in the
exponent of Eq.(10) gives correct scaling for the $<e^{\beta y_i}>$,
\begin{eqnarray}
\label{e13} <e^{y_i}>=\exp(L\phi(\beta)),
\end{eqnarray}
where $e^L$ is the number of end-points of hierarchic tree.

 As $y_i$ is a sum of
$1/\Delta v$ random variables $\epsilon$ on our tree, we define
the following distribution for the distribution of random
variables on the links $\rho(\epsilon)$:
\begin{eqnarray}
\label{e14} \rho(\epsilon)\equiv
\hat\rho(L\Delta v,\epsilon)=\nonumber\\
\frac{1}{2\pi}\;\int_{-\infty}^{\infty}\; dh \;\exp[\phi(ih)L\Delta
v-ih\epsilon]
\end{eqnarray}
For the $L\Delta v\ll 1$ we expand in the exponent the terms with
$b_l,l>1$,  and derive:
\begin{eqnarray}
\label{e15} \int d\epsilon G(x+\epsilon,v)
\rho(\epsilon)=\frac{1}{2\pi}\int_{-\infty}^{\infty}\; dh
\;\exp[b_1Ldv-ih\epsilon]\nonumber\\
\times(1+Ldv\sum_lb_l(ih)^l)G(x+\epsilon,v)= \int d\epsilon
[\delta(\epsilon-b_1L\Delta v)\nonumber\\+L\Delta v\sum_{l\ge
2}b_l(ih^l)\frac{1}{2\pi}\int_{-\infty}^{\infty}\; dh
\;\exp(-ih\epsilon)]G(x+\epsilon, v)
\end{eqnarray}
We assume a smooth behavior of G(x, v) at the the infinity $x\to \pm
\infty$, and $\frac{\partial^n G(v,x)}{(\partial x)^n}|_{x=\pm
\infty}=0$ for $n\ge 1$.

Integrating by parts we derive:
\begin{eqnarray}
\label{e16} \int d\epsilon G(x+\epsilon,v)
\rho(\epsilon)=G(x,v)+L\Delta v\phi(\frac{\partial}{\partial x})
G(x,v)
\end{eqnarray}

 Thus  we
obtain:
\begin{eqnarray}
\label{e17} G(x,v)\to [(G+L\Delta v\; \phi (\partial x)G(x,v)]^{1+L\Delta v}\approx\nonumber\\
 G(x,v)+\Delta v\; L (\phi(\partial x) G(x,v)+\; G(x,v)\ln G(x,v))\;\;\;
\end{eqnarray}
or
\begin{eqnarray}
\label{e18} \frac{\partial G(x,v)}{\partial v}=\phi (\partial
x)G(x,v)+G(x,v)\ln G(x,v)
\end{eqnarray}
where $0\le v\le 1$ plays the role of time of reaction-diffusion
equation, and we should solve the equation with the initial
distribution
\begin{eqnarray}
\label{e19} G(x,0)=\exp[-e^{-\beta x}]
\end{eqnarray}
Due to proper choose of the scaling in the exponent of Eq.(14),
there is no L dependence in Eq.(18).

 Contrary to
the KPP equation in the case of models of Ref.~\cite{de88},
Eq.~(\ref{e18}) is an exact equation.

Solving Eq.(18) for the initial distribution by Eq.(19), we can
calculate $\rho(Z)$ using the inverse Fourier transformation.

\subsection{Different distributions}

Consider different distributions, popular in cascade processes.

{\bf The normal distribution}. Taking
\begin{eqnarray}
\label{e20}\phi(k)=k^2/2
\end{eqnarray}
 we obtain
\begin{eqnarray}
\label{e21} \frac{\partial G(x,v)}{\partial v}=\frac{\partial^2
G(x,v)}{\partial x^2}+G\ln G
\end{eqnarray}
 We derived Eq. (21) in
\cite{sa02} solving the continuous branching model. First time
this equation has been derived  in \cite{do} for investigation of
non-linear diffusion.

{\bf The binomial distribution}. We have
\begin{eqnarray}
\label{e22}
\hat\rho(\epsilon)=\frac{\alpha_2\delta(\epsilon-(1-\alpha_1)}{\alpha_1+\alpha_2}
+\frac{\alpha_1\delta(\epsilon-(1+\alpha_2))}{\alpha_1+\alpha_2}
\end{eqnarray}
We get for $\phi(k)$ an expression:
\begin{eqnarray}
\label{e23}
\phi(k)=k(1-\alpha_1)+\ln[\frac{\alpha_2}{\alpha_1+\alpha_2}]+\nonumber\\
+\ln[1+\frac{\alpha_2}{\alpha_1}e^{-k(\alpha_1+\alpha_2)}]\nonumber\\
=+\sum_{n\ge
1}\frac{(-1)^n}{n!}[\frac{\alpha_2}{\alpha_1}e^{-k(\alpha_1+\alpha_2)}]^n
\end{eqnarray}
Eventually we get the following PDE:
\begin{eqnarray}
\label{e24} \frac{\partial G(x,v)}{\partial v}=G\ln
G+(1-\alpha_1)\frac{\partial G(x,v)}{\partial x}
+\nonumber\\\ln[\frac{\alpha_2}{\alpha_1+\alpha_2}] +\sum_{n\ge
1}\frac{(-1)^n}{n!}[\frac{\alpha_2}{\alpha_1}G(v,x+n(\alpha_1+\alpha_2)]
\end{eqnarray}
We are looking solutions where $G(x,v)\to 0$ for $x\to \infty$,
therefore our equation is well defined.

{\bf Gamma distribution}. Now we consider the following
distribution for positive $\epsilon$
\begin{eqnarray}
\label{e25}
\hat\rho(\epsilon)=\frac{\gamma^{\epsilon}}{\Gamma(\gamma)}e^{-\gamma
\epsilon}
\end{eqnarray}
We have the following expression for $\phi(k)$:
\begin{eqnarray}
\label{e26} \phi(k)=\gamma \ln (1+k/\gamma)
\end{eqnarray}
While we can formulate formally a PDE in this case,
\begin{eqnarray}
\label{e27} \frac{\partial G(x,v)}{\partial v}=\gamma \ln
[1+\frac{1}{\gamma}\frac{\partial}{\partial x}]\ G(x,v)+G\ln G,
\end{eqnarray}
the equation can be better formulated in Fourier space.

{\bf The Poisson distribution}. It is popular in financial
mathematics literature. We have integer values of $\epsilon$
\begin{eqnarray}
\label{e28}
\hat\rho(\epsilon)=e^{-\gamma}\gamma^{\epsilon}/\epsilon !
\end{eqnarray}
Eq.~(\ref{e9}) gives $ \phi(k)=\gamma(-1+e^{k})$. Then from
Eq.~(\ref{e18}) we derive:
\begin{eqnarray}
\label{e29} \frac{\partial G(x,v)}{\partial
v}=\gamma[-G+G(x+1)]+G\ln G
\end{eqnarray}
This renormalization group equation differs from Eq.~(\ref{e21})
in that the second order derivative is replaced by finite
difference.

{\bf Compound Poisson distribution}. The cascade processes with such
distribution are also rather popular in literature.

Now we have Poisson distribution for integers $n$ given by Eq.(28)
i.e. $p_1(n)=e^{-\gamma}{\gamma}^n/n!$ and define $\epsilon$ as a
sum of $n$ random variables $x_l$ with some random distribution
$p(x)$, defined trough the representation:
\begin{eqnarray}
\label{e30} p(x)= \frac{1}{2\pi}\;\int_{-\infty}^{\infty}\; dh
\;\exp[\alpha(ih)-ih\epsilon]
\end{eqnarray}
Thus
\begin{eqnarray}
\label{e31} \alpha(k)= \ln \int_{-\infty}^{\infty} p(y)dy\exp[ky]
\end{eqnarray}
A simple calculation gives  the function  $\phi(k)$ for the
distribution of $\epsilon$
\begin{eqnarray}
\label{e32} \phi(k)=\gamma(e^{\alpha(k)}-1)
\end{eqnarray}
Thus we get  the following PDE for $G(v,x)$:
\begin{eqnarray}
\label{e33} \frac{\partial G(x,v)}{\partial v}
=\gamma[e^{\alpha(\frac{\partial}{\partial x})}-1]G(x)]+G\ln G
\end{eqnarray}
or, using Eq.(31),
\begin{eqnarray}
\label{e34} \frac{\partial G(x,v)}{\partial v}=
 \gamma[\int dy p(y)G(x+y,v)-G(x,v)]+G\ln G
\end{eqnarray}

\subsection{The phase transition point}
{\bf General case of distribution.} The free energy of directed
polymer model Eq.(7) for any q is equivalent to the free energy of
corresponding REM with the same number of energy configurations and
distribution of $y_i$. This issue is well discussed in REM
literature \cite{de88,de90,de91,de93}.

 We have $M=e^L$ endpoints of the tree, and
any of $y_i$ has a distribution by $\rho(L,y)$, see Eq. (10). Let us
consider REM with $M$ energy level with independent distribution of
energy by Eq.(10).

At high temperature phase, assuming $<\ln Z>=\ln <Z>$, we have
\cite{de90},
\begin{eqnarray}
\label{e35} <\ln Z>=L(1+\phi(\beta))
\end{eqnarray}
 The phase transition is at the point where the entropy of free
 energy by Eq.(35) disappears \cite{de90},
\begin{eqnarray}
\label{e36} 1+\phi(\beta_c)-\beta_c\phi'(\beta_c)=0
\end{eqnarray}

{\bf Binomial distribution.} Now we have the following expression
for the free energy in the high temperature phase
\begin{eqnarray}
\label{e37} \frac{\ln
Z}{L}=\ln[\frac{\alpha_2}{\alpha_1+\alpha_2}e^{\beta(1-\alpha_1)}
+\frac{\alpha_1}{\alpha_1+\alpha_2}e^{\beta(1+\alpha_2)}]
\end{eqnarray}
Eq.(36) has no solution in this case, therefore system is always in
the high temperature phase.

{\bf Gamma distribution.} Now we have the following expression for
the free energy in the high temperature phase
\begin{eqnarray}
\label{e38}
 \frac{\ln
Z}{L}=1+\gamma (1+\frac{\beta}{\gamma})
\end{eqnarray}
Eq.(36) has no solution in this case, therefore system is always in
the high temperature phase.

{\bf The Poisson distribution}. For this case  we derive for the
free energy of high temperature phase and transition point:
\begin{eqnarray}
\label{e39}
\frac{\ln Z}{L}=1+\gamma(e^{\beta}-1);\nonumber\\
1+\gamma[e^{\beta_c}(1-\beta_c)-1]=0
\end{eqnarray}
 In the next sections we will use the travelling wave
analysis to investigate the paramagnetic phase solution of our
equations and identify the transition point. For investigation of
the spin-glass like solutions one needs in advanced mathematics of
\cite{br82}.

{\bf Compound Poisson distribution}. Now we have for the high
temperature phase the following expression for the free energy

\begin{eqnarray}
\label{e40}
\frac{\ln Z}{L}=1+\gamma(e^{\alpha(\beta)}-1)=\nonumber\\
1+\gamma (\int dx p(x)e^{\beta x}-1)
\end{eqnarray}
For the critical point we get an equation:
\begin{eqnarray}
\label{e41} 1+\gamma [\int dx p(x)e^{\beta_c x}-1-\beta \int dx
p(x)e^{\beta_c x}x]=0
\end{eqnarray}

\subsection{KPP versus new renormalization group equation}

Let us try to describe our hierarchic tree model using KPP
equation
\begin{eqnarray}
\label{e42} \frac{\partial Q(x,v)}{\partial v}=\frac{\partial^2
Q(x,v)}{\partial x^2}+Q-Q^2
\end{eqnarray}
 In \cite{de88} has been derived Eq.(42)
as an approximation of the Eq.(7) for the special case of Eq.(20).
Later Eq.(42) has been derived in \cite{ca01} as an approximate
renormalization group equation for the models of random disorder
with logarithmic correlations. While in \cite{de88,de91}  has been
already  considered the general distribution, the KPP has been
derived only for the normal distribution. In \cite{ga10}, while
using general distribution of random variables, the authors does
not used KPP. The reason is clear: for the general distribution of
random variables on the tree, the KPP equation gives wrong results
even for transition point, contrary our new renormalization group
equation which is exact.

To compare we take the case by Eqs.(28) with $\gamma=1$. The KPP
version, having the same variance of distribution
$\rho(\epsilon)$, gives
\begin{eqnarray}
\label{e43} \frac{\partial G(x,v)}{\partial v}=\frac{\partial
G(x,v)}{\partial x}+\frac{\partial^2 G(x,v)}{\partial x^2}+G-G^2
\end{eqnarray}
Using the mapping:
\begin{eqnarray}
\label{e44} G(v,x)=Q(v,x+t)
\end{eqnarray}
we return to the KPP equation by Eq.(42). This equation is well
investigated in literature. It describes a transition between
spin-glass and paramagnetic phases at
\begin{eqnarray}
\label{e45} \beta_c=\sqrt{2}
\end{eqnarray}
and free energy at paramagnetic phase is
\begin{eqnarray}
\label{e46} <\ln Z>=L(1+\beta+\beta^2)
\end{eqnarray}
 The correct transition point of the model and free energy are:
\begin{eqnarray}
\label{e47} \beta_c=1\nonumber\\
<\ln Z>=Le^{\beta}
\end{eqnarray}
Thus the KPP wrongly describes the model, and we need in new exact
renormalization group equation for the general form of
distribution of random variables on the tree.

Let us investigate our Eq.(29). $G(v,x)$ is a monotonic function
of x according to the definition by Eqs.(4),(7),
\begin{eqnarray}
\label{e48}G(v,x)\to 1,  \quad for\quad x\to \infty,\nonumber\\
G(v,x)\to 0, \quad for \quad x\to -\infty
\end{eqnarray}
Consider the case of small p in Eqs.(4),(7). The direct
calculations give
\begin{eqnarray}
\label{e49}G(v,x)=\exp[-p\exp\beta x+(v(1+\gamma(e^{\beta }-1))],
\end{eqnarray}
The latter expression is not a solution of Eq.(43). Nevertheless, it
is the solution of
 Eq.(29) when
\begin{eqnarray}
\label{e50}p\exp[\beta x+v(1+\gamma(e^{\beta }-1))]\ll 1
\end{eqnarray}

Let us investigate the behavior of the asymptotic solution for the
more general situation than the case of Eq.(50).

 Following to  \cite{de88,de91,ca01}, we assume a traveling wave like
 asymptotic
 solution
\begin{eqnarray}
\label{e51}G(v,x)=g(x+cv)
\end{eqnarray}
We get ODE:
\begin{eqnarray}
\label{e52} cg'(x)=[-g(x)+g(x+1)]+g(x)\ln g(x)
\end{eqnarray}
Eq.(52) describes a traveling wave with a front at
\begin{eqnarray}
\label{e53}x=-cv
\end{eqnarray}
We assume that for large negative x there is a solution
\begin{eqnarray}
\label{e54} g(x)=\exp[-e^{\beta x}]
\end{eqnarray}
Putting the latter expression to Eq.(52), we get the following
equation for c:
\begin{eqnarray}
\label{e55} c(\beta)=\frac{[e^{\beta}-1]\gamma+1}{\beta}
\end{eqnarray}
Eqs.(51),(54),(55) give the free energy expression by Eq.(46).

We can identify the phase transition point as the value of $\beta$
giving the maximum velocity, looking the equation:
\begin{eqnarray}
\label{e56} c'(\beta_c)=0
\end{eqnarray}
Eq.(56) gives the result of Eq.(47), $\beta_c=1$ for the case
$\gamma=1$.

At high value of $\beta$ there is another solution for $c(\beta)$,
but for its derivation one needs in advanced mathematical approach
of \cite{br82}.

Thus the investigation of our renormalization group equation,
using the qualitative approach of \cite{de88,de91,ca01}, gives
exact expression for the free energy.

\section{Random branching model}

In the previous section we derived new renormalization group
equation, using continuous deterministic branching, later
considering $q\to 1$ limit.
 Let us derive a similar renormalization group equation, using
 less abstract model of random branching \cite{de88}.

We again have a tree. It starts at some point O, and grows down,
where the vertical coordinate measures the time v. We put random
variable $\epsilon_l$ with distribution by Eq.(8) on the links
(between two adjacent nodes of the tree), replacing $\Delta v$ by
the difference of time coordinate between two nodes of the branch.
During period of time $dv$ there is a branching with a probability
dt.

 Consider the dynamics of the partition sum $Z(\beta,v)$ defined as
\begin{eqnarray}
\label{e57} Z(\beta,v)=\sum_je^{\beta y_j},
\end{eqnarray}
Here the index $j$ numerates the endpoints of our tree at the time
$v$.

 $Z(\beta,v)$ has a deterministic behavior at $v=0$:
\begin{eqnarray}
\label{e58}Z(\beta,0)=1
\end{eqnarray}
  $Z(\beta,v)$ is a random variable at $v>0$ both due to
the randomness of branching and randomness of $\epsilon_l$.

Then one has the following recursive equation for the random
partition function \cite{de88},
\begin{eqnarray}
\label{e59} Z(\beta,v+dv)=Z(\beta,v)e^{-\beta \epsilon},
\end{eqnarray}
with probability $1- dv$, and
\begin{eqnarray}
\label{e60} Z(\beta,v+dv)=(Z^1(\beta,v)+Z^2(\beta,v))e^{-\beta
\epsilon}
\end{eqnarray}
with probability $ dv$. Here $Z^1(\beta,v)$ and $Z^2(\beta,v)$ are
random independent variables with the same probability
distribution us $Z(\beta,v)$.

We identify $Z(0,v)$ with the number of endpoints of the tree, and
Eqs.(59)-(60) give the intuitive result
\begin{eqnarray}
\label{e61} <Z(0,v)>=\exp(t),
\end{eqnarray}
confirming the self-consistence of the choice Eqs.(59)-(61).
Actually the random process $Z(t)$ is defined completely trough
Eqs.(58)-(60), and we can work with these equations missing any
reference to the branching trees.

We define now, following \cite{de88}
\begin{eqnarray}
\label{e62} G(v,x)=<\exp[-e^{-\beta x}Z(v)]>,
\end{eqnarray}
Then we have an equation
\begin{eqnarray}
\label{e63} G(v+dv,x)=(1-
dv)\int d\epsilon\rho(\epsilon)G(v,x+\epsilon)+\nonumber\\
 dv G(v,x)^2
\end{eqnarray}
Eventually we get the following equation:
\begin{eqnarray}
\label{e64} \frac{\partial G(x,v)}{\partial v}=\phi (\partial
x)G(x,v)-G(x,v)(1-G(x,v))
\end{eqnarray}

For the binomial distribution we get:
\begin{eqnarray}
\label{e65} \frac{\partial G(x,v)}{\partial v}=-G(1-
G)+(1-\alpha_1)\frac{\partial G(x,v)}{\partial x}
+\nonumber\\
\ln[\frac{\alpha_2}{\alpha_1+\alpha_2}]
+\sum_n\frac{(-1)^n}{n!}[\frac{\alpha_2}{\alpha_1}G(v,x+n(\alpha_1+\alpha_2)]
\end{eqnarray}

For the Poisson distribution we get
\begin{eqnarray}
\label{e66} \frac{\partial G(x,v)}{\partial
v}=\gamma[-G+G(x+1,)]+G(x,v)^2-G(x,v)
\end{eqnarray}We can repeat the derivations of the section II-D, and again get
the same solutions for the free energy of paramagnetic phase and
the critical temperature.

 For the compound Poisson distribution we
get:
\begin{eqnarray}
\label{e67} \frac{\partial G(x,v)}{\partial v}=
 \gamma[\int dy p(y)G(x+y,v)-G(x,v)]\nonumber\\
 +G(G-1)
\end{eqnarray}

\section{Conclusions}

 We provided
an exact  (renormalization group)  partial differential equation,
for the analysis of models on hierarchic tree with continuous
branching and general distribution of random variables on the tree.
We derived new classes of reaction-diffusion equation Eq.~(18),
which sometimes has finite difference Eqs. (24),(29),(34). While
recursive equations for the hierarchic models with general
distributiqon of random variables on the tree have been well known
\cite{de88},\cite{ga10}, our article  is the first result to replace
theses recursive equation with correct PDE.

Our derivation of  renormalization-group equation is quite rigorous.
The renormalization group equation is exact, including finite size
corrections, while the renormalization group equation in \cite{ca01}
is an approximation and could be applied to calculate only bulk
characteristics of corresponding models.

The KPP equation, used in \cite{de88,de91,ca01}, correctly (exact in
the thermodynamic limit) describes the bulk free energy and critical
temperature for the hierarchic tree models with normal distribution
of random variable \cite{de88,de91}, as well as for the finite
dimensional models of disorder with logarithmic correlations
\cite{ca01}. As we checked, KPP equation failed to describe
correctly the hierarchic models with non-normal distributions. Thats
why the general case of hierarchic model, given by recursive
equation (7) derived in \cite{de88}, never has been investigated
using KPP in \cite{de88,de91}. To verify our new renormalization
group equations, we used the idea of traveling wave and methods of
\cite{de91,ca01}. Analyzing our equations, we found correct
expression for the free energy in paramagnetic phase and the phase
transition temperature.

An open problem is to construct the finite dimensional models of
disorder, which could be described by our new equations, as the
models of \cite{ca01} are described by KPP. We hope to succeed
using dynamic stochastic processes in 1 dimensional case.

Our approach (continuous branching) could be applied to
investigate the quantum disorder in d-dimensional space. In case
of continuous branching we derived exact PDE equations which
 could be solved numerically, while in alternative approach of
 \cite{ga10} has been derived approximate renormalization group
 equations and there are errors $O(1)$ in the expression of free
 energy (bulk term $\sim L$).

 The normal distribution version Eq.~(\ref{e21}) has a deep mathematical meaning:
 in some sense it describes the p-adic space with $p=1$ \cite{ko}. We hope that the investigation of these new
 reaction-diffusion equations  will  be further encouraged.

 Our renormalization group equations could also
 be applied to calculate also the complicated correlation function, as has been done in
 Ref.~\cite{sa10}
 for the string case.
This work was supported by Academia Sinica, National Science
Council in Taiwan with Grant Number NSC 99-2911-I-001-006, and
National Center for Theoretical Sciences (Taipei Branch).

\end{document}